# Energy Demand Unawareness and the Popularity of Bitcoin: Evidence from Nigeria

Moritz Platt [1,*], Stephen Ojeka [2,3], Andreea-Elena Drăgnoiu [4], Oserere Ejemen Ibelegbu [5,6], Francesco Pierangeli [7,8,9], Johannes Sedlmeir [10] and Zixin Wang [11]

[1]Department of Informatics, King's College London, 30 Aldwych, Bush House, London WC2B 4BG, United Kingdom
[2]College of Management and Social Sciences, British-Canadian University, Plot 001, Off Bebuagbong, 152102 Kakum, Obudu, Cross River, Nigeria
[3]Department of Accounting, Covenant University, Km.10 Idiroko Road, Canaanland, 112104 Ota, Ogun, Nigeria
[4]Faculty of Mathematics and Computer Science, University of Bucharest, Strada Academiei 14, 010014 Bucharest, Romania
[5]Henley Business School, University of Reading, Whiteknights, Reading RG6 6UD, United Kingdom
[6]Lagos Business School, Pan-Atlantic University, Lekki-Epe Expressway, 106104 Ajah, Lagos, Nigeria
[7]Department of Finance, University of Birmingham, 116 Edgbaston Park Road, Birmingham B15 2TY, United Kingdom
[8]Department of Political Economy, King's College London, 30 Aldwych, Bush House, London WC2B 4BG, United Kingdom
[9]Department of Economics, National University of Singapore, 1 Arts Link, Singapore 117570, Singapore
[10]Interdisciplinary Centre for Security, Reliability and Trust, University of Luxembourg, 29 Avenue John F. Kennedy, 1855 Luxembourg-Kirchberg, Luxembourg
[11]Department of Psychology and Behavioural Sciences, Zhejiang University, Yuhangtang Road, Yinquan Building, Zhejiang 310012, Hangzhou, China
*Corresponding author. Tel: +44 20 78365454; E-mail: moritz.platt@kcl.ac.uk

## Abstract

Decentralized cryptocurrency networks, notably those with high energy demand, have faced significant criticism and subsequent regulatory scrutiny. Despite these concerns, policy interventions targeting cryptocurrency operations in the pursuit of sustainability have largely been ineffective. Some were abandoned for fear of jeopardizing innovation, whereas others failed due to the highly globalized nature of blockchain systems. In search of a more effective angle for energy policy measures, this study adopts a consumer-centric perspective, examining the sentiments of Nigerian cryptocurrency users ($n = 158$) toward Bitcoin's sustainability, a representative cryptocurrency known for its high electricity demand. Three main findings emerged: 1) Even among those self-identifying as highly knowledgeable, most considerably underestimated Bitcoin's electricity consumption. 2) Participants with a more accurate understanding of Bitcoin's energy demand were more inclined to support sustainability measures. 3) Most of this supportive cohort viewed private entities as the primary stakeholders for implementing such measures. Given these findings, we suggest that consumer education should be at the forefront of policy initiatives aimed at cryptocurrency sustainability.

**Graphical Abstract**

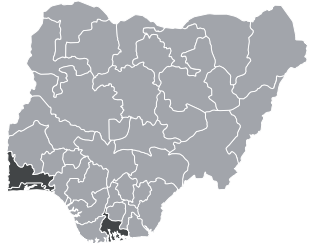

Participants predominantly resided in the *Lagos*, *Ogun*, and *Rivers* states.

**M. Platt** *et al.*

**Energy Demand Unawareness and the Popularity of Bitcoin**
Evidence from Nigeria

**Fieldwork in Nigeria: Africa's "Crypto Capital"**  *n*=158

*32% of Nigerians use cryptocurrencies.* Many as a hedge against inflation and to circumvent the limitations of an ageing banking system. We surveyed 158 Nigerian *Bitcoin* users via an online questionnaire to test the following hypotheses:

i. Most Nigerian *Bitcoin* users are unaware of its high energy demand.
ii. Users who misestimate electricity consumption see less need to counteract it.
iii. Participants who see a clear need for action feel that nongovernmental actors are responsible.

**Awareness**

Most participants (68.4%) significantly underestimated the energy demand of *Bitcoin*, while only a minority (16.5%) overestimated it.

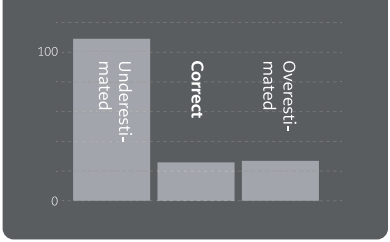

**Actionability and Responsibility**

The proportion of participants supporting measures was higher when they correctly estimated the electricity consumption (83.3%) than in the incorrect estimates group (61.9%).

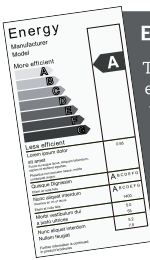

**Energy Labels as a Solution?**

The study proposes that strategies to reduce the energy demand of *proof-of-work* cryptocurrencies should target end users, promote transparency, and educate consumers. Energy labelling is suggested as a potential measure to inform users about sustainability metrics when making cryptocurrency investment decisions, drawing parallels with successful energy label initiatives in other industries.

**Lay Summary:** Cryptocurrencies like Bitcoin consume a lot of electricity, raising environmental concerns. This study surveyed 158 Nigerian cryptocurrency users to understand their awareness of the energy use of Bitcoin. The findings revealed that, although many participants considered themselves Bitcoin experts, most underestimated its energy demand. Those who were aware of the actual energy demand were more supportive of measures to reduce it. The study suggests that better educating consumers about the environmental impacts of their cryptocurrency choices, potentially through energy labelling (a practice that provides information on cryptocurrency energy efficiency to users), could lead to more sustainable practices.

# Introduction

Bitcoin is arguably the most popular cryptocurrency [1] and has a major impact on the wider crypto-asset ecosystem [2]. It has experienced enormous success since its invention in 2008, as evidenced by its peak market capitalization exceeding 1 trillion US dollars (US$) [3] and its growing user base [4]. Yet, this original 'pure digital asset' [5] has been criticized outright for reproducing societal inequality [6] and for its enormous electricity demand caused by the underlying proof-of-work (PoW) consensus mechanism [7–10]. Although Bitcoin's exact energy footprint cannot be established with certainty and estimates vary considerably depending on the method of measurement applied [11, 12], the cryptocurrency is commonly considered a substantial contributor to global warming. As such, it was found to produce up to 65.4 Mt $CO_2$ annually: the equivalent of the total emissions of Greece [13]. There are now numerous cryptocurrencies that incorporate more sustainable consensus mechanisms [14], including the second largest cryptocurrency by market capitalization, Ethereum, which has recently transitioned to proof-of-stake (PoS) [15]. Still, many cryptocurrencies, first and foremost Bitcoin, continue to apply PoW. Consequently, a critical examination of this technology remains pressing beyond the ongoing debate on cryptocurrencies as instruments of payment in the context of criminal activity [16, 17].

Although many experts, including many in the wider Bitcoin community, continue to emphasise that the security of the tried and tested PoW mechanism is unrivalled [18–20], this view is not shared universally [21–23]. The strengths and weaknesses of PoW and PoS from the perspective of economic security and decentralization remain the subject of debates [24]. Some publications address misconceptions concerning the electricity consumption characteristics of blockchain applications in general [12] and PoW cryptocurrencies in particular [25]. However, we have no knowledge of academic works that directly assess the awareness of electricity consumption of cryptocurrency users.

## Research Objectives

This research seeks to bridge the gap between cryptocurrency understanding and environmental consciousness within the Nigerian context. As the global conversation around Bitcoin's environmental impact intensifies and with the rising prominence of cryptocurrency transactions in Nigeria, the insights from this study can potentially shape policies, steer educational initiatives, and guide stakeholder decisions.

### Research Gap and Study Context
Despite studies on the sustainability of payment systems by regulatory bodies [26, 27] and various attempts to regulate cryptocurrencies [28, 29], legislators still lack an understanding of how users perceive policies targeting cryptocurrencies and whether they have the knowledge needed to understand the motivation behind these policies. Due to the decentralized nature of cryptocurrencies and the corresponding challenges in banning PoW-based cryptocurrencies, measures that consider users' perspectives and actions are likely to be the only way to bring about long-term improvements in energy use. Considering that the cryptocurrency market offers products with dramatically different carbon footprints, there is a clear gap in research on potential mechanisms to encourage users to explore more sustainable alternatives.

This paper aims to address this gap through field research in Nigeria. Nigeria's social, cultural, and economic realities are particularly conducive to such research: as a result of a recession and rising inflation, the Nigerian economy is experiencing stress [30]. Coupled with the deficiencies of an outdated and costly traditional banking sector [31], this implies that many Nigerians regularly and routinely use cryptocurrencies as a means of payment [32] despite the rejective position of the government [33]. This application stands in stark contrast to industrial countries where cryptocurrencies are pursued primarily as a speculative form of investment [34–37]. Furthermore, the Nigerian public is aware that, due to their location, they could be severely affected by the effects of climate change [38–40]. The combination of the extensive use of cryptocurrencies and the high awareness of the effects of climate change makes Nigeria a suitable setting for our study.

### Hypotheses
We collect and quantitatively analyse data from 158 cryptocurrency users in Nigeria, focusing on hypotheses in three areas: awareness, actionability, and responsibility.

#### Awareness
The starting hypothesis of this work is that most Nigerian Bitcoin users are unaware of its high energy demand. This hypothesis is motivated by broader research on the technological awareness of Bitcoin users in similar markets [41]. It is furthermore influenced by earlier findings that showed that cryptocurrency users did not take sustainability into account when selecting a cryptocurrency to use or mine [42].

#### Actionability
It can be further hypothesized that users who misestimate electricity consumption see less need to counteract it. This is conceivable because a correct understanding of the causes of global warming was found to be a key determinant of behavioural intentions to act against it [43]. Furthermore, it can be speculated that those users who can accurately estimate electricity consumption possess sufficient expertise to contemplate countermeasures.

#### Responsibility
A further hypothesis is that participants who see a clear need for action counteracting the electricity consumption of Bitcoin feel that nongovernmental actors are responsible because distrust in government has been found to be linked with cryptocurrency adoption [44].

### Outline
In section 'Background' of this paper, we introduce the fundamentals of blockchain technology, covering technical concepts like consensus mechanisms and their application to cryptocurrencies like Bitcoin. In this context, we also describe the drivers of electricity consumption in PoW cryptocurrencies. Subsequently, we present critical insights into cryptocurrency adoption in Nigeria and explain the situation in China as a case study. Next, in section 'Related Work' we give an overview of related work with a focus on research on cryptocurrency user attitudes. The remainder of this paper features a description of the questionnaire-based online survey conducted (see section 'Method'), followed by a discussion of the results obtained (see section 'Results'). We conclude with policy considerations and avenues for future work (see section 'Conclusions').

## Background
Blockchain technology, a kind of distributed ledger technology (DLT), goes back to the work of Nakamoto [45] and forms the foundation of the most common cryptocurrencies [46], including Bitcoin. A 'blockchain' is commonly characterized as a linear, append-only collection of data elements ('blocks'), all of which are linked to form a tamper-evident chain using hash pointers [47, 48]. The data in blocks can be arbitrary [49] but, in the case of blockchains with native cryptocurrencies, consist mainly of transfer instructions between accounts [50]. Although, in theory, transfers can occur between arbitrary addresses, real-world systems typically constitute small-world networks [51]. Different entities hold replicas of chains and synchronize them by means of a consensus mechanism that facilitates decentralized agreement on which data elements to append next [52]. Often, blockchains that expose a native cryptocurrency provide incentives to those users who participate in consensus [53]. In the context of PoW, the incentives are called 'mining rewards'. Stakeholders who aim to transfer amounts of cryptocurrency offer transaction proposals to block producers, who, in turn, select transactions to maximize their reward in terms of fees [54].

Blockchain technology constitutes the foundation for most existing cryptocurrencies. This is because blockchain technology is well suited to record account balances in decentral systems that allow anyone to participate, yet do not require a distinguished trusted authority [55].

### Cryptocurrency Mining
As noted, blockchain technology aims to provide a decentralized ledger that is synchronized across distributed replicas. To provide synchronization that is not dependent on the availability and honesty of a distinguished entity, a wide variety of consensus mechanisms can be applied [56]. These typically combine economic incentives and cryptographic protocols to achieve a system state in which all honest nodes come to agreement under the assumption of an honest majority of nodes [25, 57]. Initial research on consensus mechanisms dates back to the 1980s with the work of Lamport [58] on 'Paxos' and Castro and Liskov [59] on Practical Byzantine Fault Tolerance as key contributions.

Early work focused on 'closed' systems in which the number and identity of participating parties are determined in advance. For example, in an aeroplane that requires a particularly high reliability of sensor information, it may be of interest that a coherent overall picture of the system state can be formed even if some components behave unpredictably, for instance, in the presence of cosmic radiation. In such closed systems, the problem of consensus can be solved efficiently by majority voting combined with appropriate communication protocols.

In contrast, 'open' systems, such as many cryptocurrencies, do not have predetermined groups of users. Consequently, they do not conform to the principle of 'one participant, one vote' [25]. In such systems, an entity that intends to control the system could skew majority votes by registering a large number of bogus accounts, a technique known as 'Sybil attack' [60]. Most commonly, preventing such attacks is done by linearly tying the weight of a vote to a scarce resource provided by the participants that is verifiable digitally and by encouraging the provision of this resource through economic incentives [25]. The earliest and, arguably, simplest approach to satisfy this requirement is to use computational power as a scarce resource, as first proposed by Dwork and Naor [61] and later applied by Nakamoto [45] in the context of the consensus mechanism for the first cryptocurrency Bitcoin. This approach, commonly termed proof-of-work (PoW), ultimately ties a voting weight to hardware and energy and, thus, to capital. More precisely, miners compete by solving cryptographically hard puzzles through trial and error [62]. Whoever solves a puzzle can submit its solution along with the transactions collected as a new block, which will be accepted by other honest nodes. Next, all honest nodes aim to find a subsequent block, including a corresponding solution to a puzzle that is linked to the previous block.

### Electricity Demand

As a consequence, the electricity demand of a PoW cryptocurrency can be determined via a simple approximation: assuming participating miners are rational, they will only provide computational power if their expected revenue (i.e. rewards for finding new blocks and the fees of the transactions included in it) exceeds the cost that they incur for buying, maintaining, and operating hardware. At the time of writing, Bitcoin releases a reward of 6.25 Bitcoin (BTC) for creating a block, and on average, producing a block takes 10 minutes [63]. Cumulative transaction fees have consistently been one to two orders of magnitude lower per block than mining rewards in many cryptocurrencies, including Bitcoin [25] and, therefore, can be ignored for rough estimates. Following this line of reasoning, a simple worst case model for electricity consumption can be established by assuming electricity costs are the only costs for miners and a lower bound on electricity prices is 0.05 US\$ per kW h [25, 63]. The accuracy of this model can be improved by considering that the share of electricity costs in mining is only ~40% [8]. In general, the decline in hash rate after price shocks on the revenue side (e.g. a sharp decrease of the Bitcoin price and halving events) and on the cost side (e.g. a sharp increase in electricity prices at the end of the rainy season in China) suggest that the upper bound is relatively accurate [64, 65]. On the other hand, a lower bound for electricity consumption can be derived by observing the complexity of the solved puzzles and the distribution of the energy efficiency of the mining hardware deployed. A variation of this method is also applied by the Cambridge Bitcoin Electricity Consumption Index (CBECI) Bitcoin network power demand model[1], a widely recognized consumption model [67–68], which forms the basis of the consumption figures applied in the survey (see subsection 'Materials'). As of mid-July 2022, Bitcoin's annual electricity consumption is within the theoretical limits of 40–138 TW h, with an estimate of 84 TW h according to the CBECI model.

This number seems enormous on its own, yet criticism ignites even further when considering the energy requirements per transaction because cryptocurrencies only have very limited transaction processing capacity [69]. For instance, Bitcoin processes around four transactions per second; the theoretical maximum (given the currently accepted system parameters) is around seven transactions per second [70]. Mathematically this yields ~660 kW h per transaction, more than an average household in Germany consumes in 2.5 months, or as much as the average annual electricity consumption of four Nigerians. Nonetheless, it is important to note that because transaction fees currently play only a marginal role in the remuneration of miners, increasing the limit of transactions in the Bitcoin protocol would not increase total electricity consumption considerably. Due to this particularity, the 'energy per transaction' metric frequently causes misunderstandings [12, 25, 71, 72]. In this survey, we will therefore consider annual electricity consumption as described in subsection 'Network-Wide Electricity Consumption as Anchor Point'.

The economic relationship between the total electricity consumption of a PoW-based blockchain and the market price of its native cryptocurrency suggests that increasing the energy efficiency of the mining hardware or reducing the number of transactions will not help much to improve sustainability [25]. Because both mining rewards and transaction fees are paid in the native cryptocurrency (e.g. BTC) of the blockchain, the cryptocurrency market price is the most important factor influencing blockchain electricity consumption systematically [11]. This price, however, cannot be directly controlled through regulatory means. Instead, a reduction in adoption of a given currency is likely to reduce its price [73] and can therefore be considered a sustainability policy tool. For this reason, we focus on evaluating whether consumers who are well informed about the electricity consumption of a PoW blockchain and its drivers are more willing to support countermeasures such as abandoning an unsustainable blockchain and moving to a sustainable one.

### Alternatives to PoW

A popular alternative consensus mechanism used in cryptocurrencies is PoS, which uses cryptocurrency stake as a scarce resource instead of computational work. As we illustrate in Fig. 1, systems that rely on this consensus mechanism are several orders of magnitude more energy-efficient than those that use PoW [22, 74]. Although PoS-based systems are arguably more challenging to design and implement securely and there are doubts about their incentive compatibility under certain conditions [18, 19], the question of whether PoW or PoS is more secure remains unresolved [75, 76]. In any case, PoS-based systems enable consumer choice: although end users are challenged by the usability of blockchains in general [77], whether systems are operated using PoW or PoS does not noticeably affect their usability [78]. Energy-intensive PoW-based currencies, foremost Bitcoin, remain dominant [79], although more sustainable PoS-based alternatives exist and are conveniently available to users through popular centralized exchange websites [80] that offer comparatively good usability and low transaction fees [81].

### Cryptocurrency in Nigeria

Nigeria, distinguished by a high position in the 'Bitcoin Market Potential Index' [82] and considered an attractive environment for commercialization of cryptocurrency activities [83], has a cryptocurrency usership of ~32% of the population [32, 84].

---

[1] See https://ccaf.io/cbeci/index.

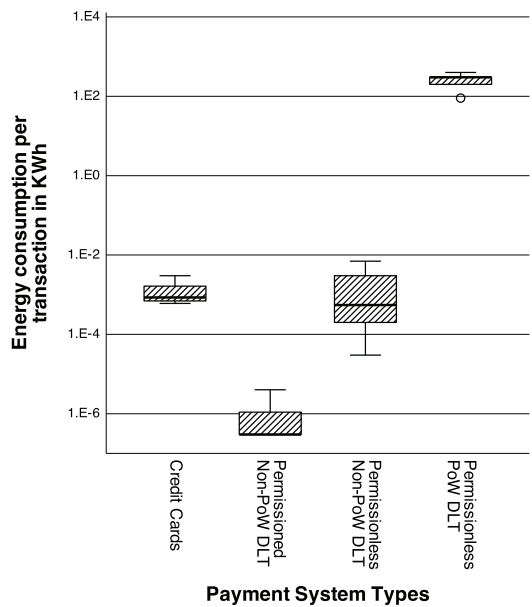

**Figure 1.** A comparison of data concerning the electricity consumption per transaction reported in the wider literature (logarithmic scale) compiled by Agur et al. [26] confirms that the scientific community, despite divergent estimates, considers permissionless PoW DLT systems to consume several orders of magnitude more electricity than other payment systems.

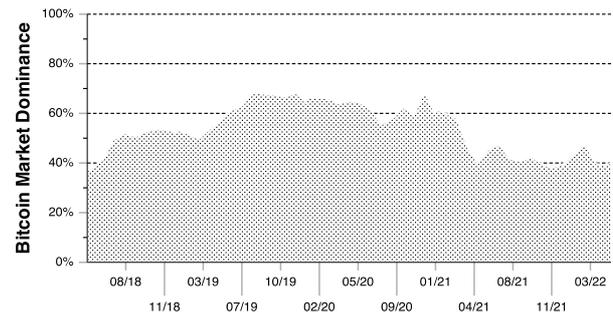

**Figure 2.** Bitcoin's market capitalization relative to that of all other cryptocurrencies combined between 2018 and 2022 [79].

This high number may be explained by economic hardship associated with the prevailing unemployment, which is believed to be structural [85], coupled with a worsening inflation [32]. This climate, linked with the proliferation of mobile and wireless devices [86], allowed many citizens, especially the youth, to interact with and adopt cryptocurrencies as a safe haven from looming inflation [87]. Furthermore, high fees for international transfers of funds have established cryptocurrencies as an alternative to traditional banking [88]. Consequently, Nigerians consider international acceptance as one of the key advantages of cryptocurrencies [89]. In addition to legitimate applications, cryptocurrencies have been found to be used for illicit purposes in Nigeria, including money laundering [90] and financing acts of terrorism [91]. Furthermore, they can be used in the context of scams around fabricated cryptocurrencies or cryptocurrency theft [92], activities that entrenched youth unemployment contribute to [93]. Nigeria has become infamous for such scams [94–96], and cryptocurrency transactions, due to their non-reversible nature, hold a special allure for scammers [97]. Despite these concerns, cryptocurrency activity in Nigeria is believed to predominantly help people address their daily financial needs [98], contrary to the portrayals in popular media. This sentiment was echoed by many Nigerians we interacted with in the course of our fieldwork.

Due to the rising popularity of cryptocurrencies, concerns arose among the regulatory authorities, especially the Central Bank of Nigeria (CBN): cryptocurrencies were seen as excessively speculative in nature and therefore considered a risk to the financial well-being of Nigerians [3, 99, 100]. Therefore, in an effort to regulate the market, the CBN placed a ban on banks that facilitate cryptocurrency-related transactions in 2017 [101]. This, however, remained largely unenforced [102]. In another swift move by the CBN, after the initial order was dropped in 2021, an initiative was taken to protect the public and safeguard the country from potential threats posed by 'unknown and unregulated entities' that are 'well-suited for conducting many illegal activities' [100]. In this context, the CBN directed banks to stop using their platforms to transact or engage with entities that are involved in cryptocurrency activity [101, 103]. In addition, they were asked to close accounts of individuals and institutions involved in cryptocurrency transactions [100]. In April 2021, three banks were sanctioned with an 800 million Nigerian naira (NGN) (∼2.1 million US$) fine for failing to prevent customers from engaging in cryptocurrency transactions [104]. Since then, many Nigerians have reported that their bank accounts have been frozen due to cryptocurrency-related activity. Approximately the same time, the CBN launched a project to improve the efficiency of payment systems [105] by implementing a centrally issued and regulated Central Bank Digital Currency (CBDC), which is, at the time of writing, the only fully adopted CBDC on the African continent [106]. The resulting system, however, notably lacks decentralization and decoupling from the fluctuation of the NGN and was therefore not widely recognized as a replacement for cryptocurrencies [107]. Therefore, and because it lacked other desirable characteristics such as interest-bearing properties or feelessness [108], ultimately, this CBDC achieved only 'disappointingly low' public adoption [109]. Despite this initiative and the legislative focus on cryptocurrencies, many citizens remain highly committed to them.

## Regulating Cryptocurrency Activity

Bitcoin is the first and arguably one of the most relevant applications of blockchain technology. As such, it can be considered the archetype for cryptocurrencies [110] because it served as inspiration for most of the large number of alternative systems in the space [111], including some that are directly derived from its core protocol (e.g. 'Litecoin'). After more than a decade, Bitcoin still accounts for around half of the cryptocurrency market capitalization [112], a condition that is known as 'Bitcoin dominance'. Fig. 2 shows how this measure has fluctuated in recent years with low points <40% in 2018 or the second half of 2021 due to the rise of 'altcoins', alternative digital currencies, to peaks >70% in times of uncertainty and market volatility.

Bitcoin's dominance may reflect the widespread and growing adoption that ultimately led it to be among the best performing assets of the last decade, outperforming many stocks, bonds, commodities, and traditional currencies [113]. There are now proposals at the institutional level to allow banks to keep 1% of reserves in Bitcoin [114]. Thus, this study focuses on users of this archetypal cryptocurrency.

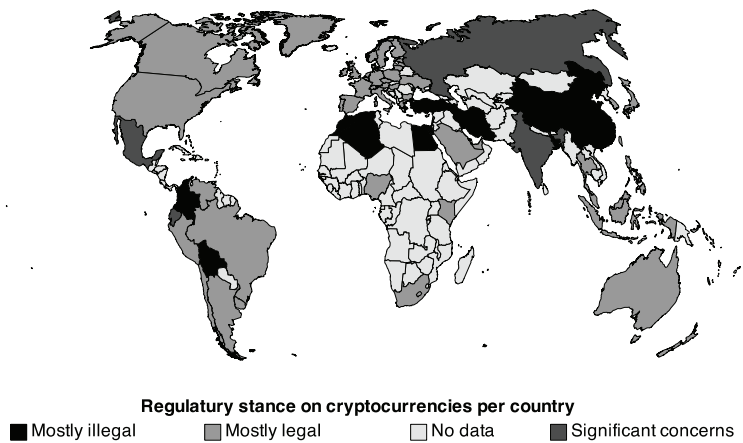

**Figure 3.** The visualization of the global regulatory stance on cryptocurrencies adapted from the work of Hammond and Ehret [125] shows that cryptocurrencies are considered mostly illegal by the governments of Algeria, Bangladesh, Bolivia, China, Colombia, Egypt, Iran, Morocco, and Turkey.

Despite being remarkably successful, cryptocurrencies have encountered numerous setbacks: contentious issues span from facilitating money laundering [115, 116] to concerns regarding their impact on the environment [117, 118]. Although this has sparked regulatory interest, decentralized and transnational cryptocurrency systems challenge more traditional regulatory sandboxing [119]. Consequently, relatively little validation of the regulatory compliance of the processes and practices surrounding cryptocurrencies has been undertaken [120]. This has contributed to regulatory gaps and thus, to legal uncertainty [121]. This affects not only environments with conservative attitudes toward digitalization but also jurisdictions that are generally perceived and seen as leaders in such matters, for instance, South Korea [122] or the United Arab Emirates [123, 124].

A plethora of different approaches, mostly founded in theory, have been observed in recent years. Some regulators allowed experimentation and showed tolerance; others opted for implicit or absolute bans (see Fig. 3). Numerous topical academic works considered regulatory aspects [126]: regulatory measures in the past focused, for instance, on fiscal interventions addressing miners [117, 127, 128], an approach with relevance beyond cryptocurrencies as evidenced by the increasing attention environmental taxation receives [129]. Some proposed measures revolved around introducing sustainability criteria for institutional financial market actors [130] or on prohibitive regulations concerning miners [29, 131]. Furthermore, design-side policies, such as pushing for voluntary redesigns of PoW protocols, were proposed [29]. Cryptocurrency developer communities, however, apply decentralized governance and exhibit autonomous characteristics [132]. Therefore, design-side policies are likely to suffer from a lack of enforceability. Consumer-focused policies are rarely proposed, and where they are, often make unrealistic assumptions, such as sovereign control over internet traffic [133]. Regulating cryptocurrencies remains challenging [134], and policymakers seem to consistently underestimate the technical complexity involved in efficiently targeting this novel phenomenon [135].

### The Chinese Example

A case study that illustrates these regulatory challenges is China: in 2017, Chinese authorities started to severely restrict the use of digital currency because the government was concerned they were facilitating capital outflows as well as money laundering and other fraudulent activities. This led to banning initial coin offerings—cryptocurrency-based avenues to raise funds via the issuance of digital tokens and without the participation of a trusted and regulated authority [136]. Second, China passed regulations to prohibit exchanges of BTC and Renminbi (RMB), effectively cutting the link between traditional financial intermediaries and cryptocurrency markets. At the same time, the People's Bank of China, the country's central bank, issued several warnings concerning Bitcoin and other cryptocurrencies, reminding the public that these do not enjoy the same legal status as fiat currencies. This had a major impact on BTC–RMB trading volume and caused spillover effects to geographically close regions shortly after the introduction of more restrictive regulations, including an increase of >25% and 20% in the trading volume of Bitcoin in Korean won and Japanese yen, respectively [137]. Chinese peer-to-peer exchanges, where buyers and sellers are matched directly, also registered ample trading increases because they provided a way to bypass regulation.

More recently, China has adopted an even more rejective stance on cryptocurrency-related activities, particularly Bitcoin mining. In 2019, the government termed Bitcoin mining 'undesirable', a label used for industries that should be restricted or phased out by local governments. Finally, in 2021, Bitcoin mining was effectively forced to shut down due to environmental concerns brought forward by the government. As illustrated in Fig. 4, the ban was initially effective: mining activity halted in the summer of 2021, with China's hash rate, a measure of mining speed in PoW, going to zero. The two countries that benefited most from the ban in terms of share of the global hash rate were the USA and Kazakhstan.

However, given the significant mining capability built in previous years, which at the time accounted for >70% of the global share, some Chinese miners were able to return to their activities despite bans, making China the country with the second-largest share of Bitcoin mining activity globally. Others moved hardware to less strictly regulated regions and continued activities there. Whether or not miners will continue to elude the ban in the future is unclear, but the events outlined suggest that even for countries with a strong grip on economic activities, such as China, enforcing outright bans is highly challenging. This perspective is consistent with work by Chen and Liu [138], who investigated the impact of government Bitcoin trading interventions on the activities of Chinese investors. They found that, although Chinese participation in the Bitcoin market has decreased, local actors remain deeply involved. Ultimately, the Chinese example shows that it is not sufficient for regulators to ban and discredit cryptocurrencies to effectively prevent adoption [139].

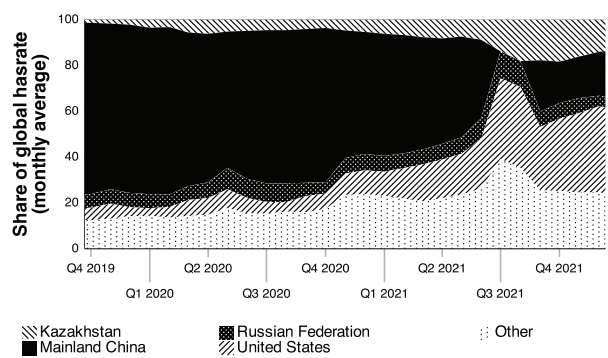

**Figure 4.** The evolution of mining activity between 2019 and 2022 (reproduced from CBECI data).

## Related Work

We consider the following types of studies as relevant related work: first, articles that target cryptocurrency users beyond Nigeria through survey research (see subsection 'Cryptocurrency User Attitudes'), thus exposing patterns of user attitudes, behaviours, and experiences. Second, articles that investigate issues related to Bitcoin in the Nigerian context (see subsection 'Cryptocurrency in Nigeria') from legal, regulatory, or macroeconomic perspectives.

### Search Strategy

We used the database *Web of Science*, which is widely acknowledged in the librarian and research communities for listing highly relevant peer-reviewed content [140]. Initially, we ran the search query (`cryptocurrency OR cryptocurrencies OR bitcoin`) AND `attitude` to retrieve works on cryptocurrency user attitudes (see subsection 'Cryptocurrency User Attitudes'). Subsequently, we ran the query (`cryptocurrency OR cryptocurrencies OR bitcoin`) AND (`nigeria`) to obtain relevant literature on cryptocurrencies in the Nigerian context (see subsection 'Cryptocurrency in Nigeria'). Finally, we manually screened the abstracts of the identified manuscripts for relevance to our work and included those we deemed relevant in the corresponding subsections. We also added references to some works as a consequence of addressing reviewer comments.

### Cryptocurrency User Attitudes

Most of the relevant prevous literature of which we are aware examines predictors of interest in using cryptocurrencies. A notable number of studies apply established psychological techniques, such as the Theory of Planned Behaviour (TOPB) [141], one of the most widely used theories of behavioural prediction. Some studies investigate which predictors influence the willingness to adopt cryptocurrencies 'in general' [142–146]. Other studies identify predictors that influence users' decision to consider them as a 'form of investment' in particular [147, 148]. Research has been conducted on the circumstances under which users tend to support cryptocurrency as a means of payment by Kim [149] and Salcedo and Gupta [150]. Due to the different foci of these surveys and the variety of methods used, the studies come to diverse, and, at times, contradictory conclusions.

For instance, Bashir et al. [151] find that gender and social circle are decisive factors for Bitcoin ownership. Mazambani and Mutambara [145], Schaupp and Festa [146], and Pham et al. [147] conclude that attitudes toward the behaviour of using cryptocurrencies are the determining construct in the context of TOPB. Steinmetz et al. [37] conclude that German cryptocurrency users are predominantly young, male, well-educated, and affluent. Gagarina et al. [152] confirm the common belief that a liberal worldview correlates with the intention to use cryptocurrencies [153]. Seemingly in conflict with this are the findings of Albayati et al. [143], whose results suggest that users are more interested in adopting cryptocurrencies when their activities are regulated and secured by the government. Alaklabi and Kang [142] conclude that technological awareness has a positive influence on the intention to use cryptocurrencies. This is consistent with the finding of Smutny et al. [148] that shows that a lack of information on the operating environment is a disincentive to cryptocurrency investment. Anser et al. [144] show that a high level of activity in social media correlates with the willingness to use cryptocurrencies. The findings of Kim [149] similarly present a picture of cryptocurrency users focused on social presence: the dimension 'power–prestige' was established as the most influential factor in the approval of Bitcoin. Salcedo and Gupta [150] argue that cultural values and norms have a major impact on the willingness to use cryptocurrencies: collectivists, as well as representatives of long-term-oriented cultures, were found to be inclined toward blockchain technology.

In summary, the existing literature portrays users of cryptocurrencies as maintaining interpersonal relationships with their peers, being technically savvy, well informed, well networked, and having libertarian worldviews. This user group is also strongly represented in our work. However, the attitude of this group toward sustainability has not played a significant role in the scientific discourse so far.

### Cryptocurrency in Nigeria

Academic coverage of issues related to cryptocurrencies in the Nigerian context is sparse. Many previous works position Bitcoin as a technology discovered by Nigerians in the context of the 2016 recession as a stable alternative to the rapidly depreciating Naira [154]. To our knowledge, only two quantitative academic studies based on questionnaires have been conducted with a focus on Bitcoin in the Nigerian context: Eigbe [155] investigates the level of awareness and adoption of Bitcoin in Nigeria, finding that most of the respondents lacked a proper understanding of the functionalities of Bitcoin, even if they claimed otherwise. A study by Salawu and Moloi [156] targets Nigerian professional accountants: they were considering offering services in a cryptocurrency environment, although most indicated that the enactment of specific legislation would be a prerequisite for doing so.

The prevailing sentiment throughout the relevant works is that Bitcoin in the Nigerian context is not a passing fad but is of significant societal importance. This is reflected in a study by Jimoh and Benjamin [157] that underlines the macroeconomic importance of Bitcoin by showing that the volatility of cryptocurrency returns has a measurable impact on the broader financial markets in Nigeria. Egbo and Ezeaku [158] underscore the serious disruptive potential of cryptocurrencies by showing that these are threatening the very foundation of the business of commercial banks operating as intermediaries in Nigeria. Whereas the previously outlined works highlight the potentially positive impact of cryptocurrencies on the Nigerian economy, other works focus on negative aspects, such as the risks of using cryptocurrencies for the financing of terrorism [91], negative effects of cryptocurrencies on the exchange rate [159], or the inability of Nigerian legislation to effectively target cryptocurrency-related activities [160, 161].

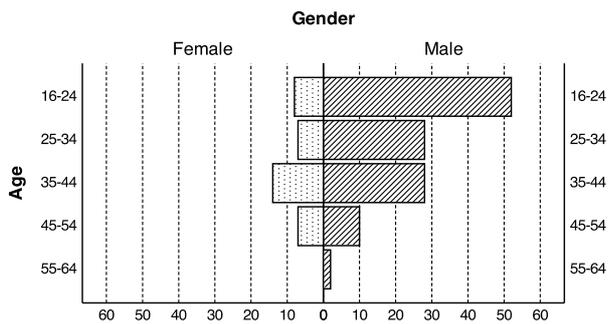

**Figure 5.** Absolute frequencies of distribution of valid responses for gender and age dimension.

## Method

We designed our study as a questionnaire-based online survey. To minimize the risk of data quality issues, a local research data collection provider was tasked with collecting data by individually approaching potential participants and ensuring that they were members of the target population. Owing to the unclear regulatory situation in Nigeria and the fear of legal repercussions that may arise from it, recruiting participants proved challenging, but was ultimately successful because most participants felt reassured about their anonymity. This was helped by the fact that the study was led by a UK institution, as opposed to a local one.

## Participants

One hundred fifty-eight valid responses were collected between 25 November 2021 and 30 March 2022 by convenience sampling. All participants were aged ≥16 years and resided in Nigeria. All participants reported having undertaken at least one Bitcoin transaction in the last 5 years at the time of this study.

Ethics approval was obtained before participant recruitment began. Participant recruitment had two avenues. First, Covenant University students from Ota, Nigeria, who had a verified interest and background in cryptocurrency, as evidenced by extracurricular activities, were approached and offered opportunities to participate voluntarily. Second, the study was advertised in Nigerian cryptocurrency groups on the Telegram messenger, the *de facto* messaging platform for the cryptocurrency community [162]. For both approaches, the participants were self-selected and did not receive compensation. The sample obtained is biased toward male participants, with 76.9% of the respondents identifying as male (see Fig. 5). This imbalance may be traced back to the convenience sampling method in conjunction with a more pronounced interest in cryptocurrencies as investment instruments among younger men [163, 164].

## Materials

The questionnaire contained a total of 107 items on eight pages, some of which were conditional[2]. It used screening questions throughout the survey to ensure that only members of the target population participated.

---

[2] See supplementary material for the original questionnaire.

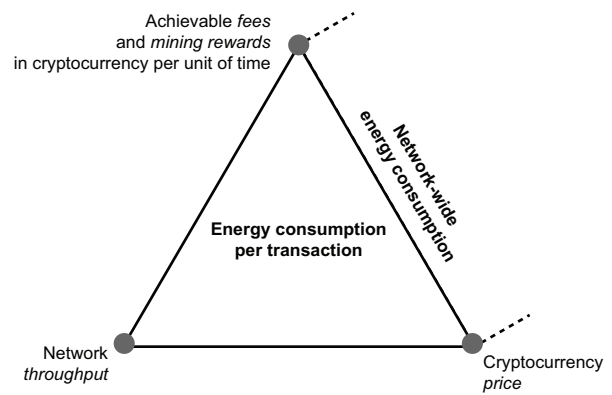

**Figure 6.** Comparison of network-wide electricity consumption models and transaction-based consumption models.

The questionnaire was designed to measure the degree of expertise in cryptocurrency technology participants possess. It was furthermore designed to measure how accurately participants estimate Bitcoin's electricity consumption. Finally, it measured the degree to which participants believe that Bitcoin's electricity consumption poses an environmental issue, whether measures should be taken, and which stakeholders, if any, they consider responsible for acting against it.

### Network-Wide Electricity Consumption as Anchor Point

As described in subsection 'Electricity Demand' and visualized in Fig. 6, typically, PoW electricity consumption is quantified either on a system-wide basis (taking into account transaction fees, block rewards available to miners, and the price of the cryptocurrency) or per transaction by additionally considering network throughput.

Consequently, when preparing the materials for this survey, the choice arose as to whether users should be questioned about their assessment of electricity consumption per transaction or about network-wide electricity consumption. We decided on the latter, considering that an increase in the transaction throughput of Bitcoin does not cause a substantial increase in its total electricity consumption. As such, a user who decides to engage in a Bitcoin transaction will not directly contribute to increasing the electricity consumption of the system but only via secondary effects, such as increases in the transaction fee levels and cryptocurrency prices, owing to increased popularity. Yet, in our experience, most users (and even researchers [165]) are not aware of this nuance and assume that additional transactions will proportionately increase Bitcoin's total electricity consumption. Therefore, it seems more appropriate to survey users about network-wide electricity consumption. In Q21, participants were, therefore, presented with the following question to which six potential answers were provided (see Table 1):

> What do you estimate the electricity requirements of operating the entire Bitcoin network to be?

We assume a value of 121.46 TW h to be close to the actual annual electricity consumption of Bitcoin at the time of conducting the survey. This value is the median of daily estimates of annualized consumption[3] during the data collection period. Estimates fluctuated between 108.08 and 140.11 TW h during data collection.

---

[3] See https://ccaf.io/cbeci/index.

**Table 1.** We asked questions to gauge how realistically participants estimate the annual electricity consumption of Bitcoin. We wrote the questions so that the participants could relate electricity consumption to the realities of their lives. The median of estimates obtained from the CBECI is printed as 'actual' value for the benefit of the reader only and was not shown to the participants

| Question Wording | Annual consumption (TW h) |
| --- | --- |
| Similar to the overall electricity consumption of a small town in Nigeria | 0.0006 |
| Similar to the overall electricity consumption of the Lagos Metropolitan Area | 5.8 |
| Similar to the overall electricity consumption of Nigeria | 29 |
| **Actual**: *Total Bitcoin electricity consumption (CBECI)* | 121.46 |
| About four times the overall electricity consumption of Nigeria | 116 |
| Similar to the overall electricity consumption of the entirety of the African continent | 700 |
| Similar to the overall electricity consumption of the entirety of the African and European continents combined | 2845 |

This shows that, during the survey period, option four (about four times the overall electricity consumption of Nigeria) was the most accurate estimate on all days.

We offer a wide range of potential answers that correspond to electricity consumption figures between 600 MW h and 2,845 TW h. The values chosen as potential answers were deliberately extreme to avoid ambiguity. Because participants are unlikely to have a reference point for physical units of measurement, the electricity values were not exposed in the questionnaire. Instead, examples that are relatable to the participants' living situation (see Table 1) were used.

*Experience Assessment and Opinions*

Some sections of the questionnaire reuse parts of existing surveys. This is also the case in Q19, in which we ask participants for their reasons for acquiring cryptocurrency:

> Why have you acquired Bitcoin in the past?

This question, along with others in the cryptocurrency experience section of the survey, reproduced questions from the *OECD Consumer Insights Survey on Cryptoassets*, a questionnaire that 'has been designed to survey consumers/retail investors to collect data on their attitudes, behaviours, and experiences toward digital financial assets, specifically digital (or crypto) currencies and initial coin offerings' [166].

Subsequent parts of the questionnaire assess the degree of concern participants have regarding some effects of climate change. For example, Q20 asks about participants' areas of concern in the context of climate change:

> How concerned are you about the potential consequences of climate change to your living environment?

The options presented were taken from previous work by Haider [39], who summarizes the likely impacts of climate change in Nigeria based on previous studies. By using this previous work, the options presented to participants were tailored to the effects that were most likely to affect them.

To ensure the quality of measurements, the local research data collection provider conducted a pilot study with 12 participants, assessing the understandability of research materials with members of the target population before the commencement of data collection. Some changes to the survey materials were implemented according to the findings of the pilot study and thereby improved comprehension of the materials in the target population.

## Procedure

Of 1,088 participants that started the survey, 158 completed it. Participants completed the questionnaire within 19 min 47 s on average. Deception was not used during study recruitment. Participants were told that the study was designed to understand their attitudes toward cryptocurrency use and environmental issues. They were then informed about the fact that their participation is completely voluntary and that they should only take part if they want to. Furthermore, they were educated about the fact that choosing not to participate would not disadvantage them in any way. The research data collection provider then made them aware that they would be provided with an information sheet for participants before answering any questions. Those persons that expressed an interest in participating after this introduction by the research data collection provider were given a survey link, either in the form of a printout, via e-mail, or via Telegram message. The research data collection provider had no knowledge of whether the potential participants indeed followed the link.

Some participants raised a serious concern that their identity could be revealed to Nigerian authorities. This concern stemmed from the fear of facing legal repercussions by the Nigerian government, which has taken a rejective stance toward cryptocurrencies (see subsection 'Cryptocurrency in Nigeria'). The research data collection provider was able to alleviate some of the concerns by pointing to the applicability of the UK General Data Protection Regulation; however, some potential participants were not convinced by this argument and remained disinterested in participation.

Once participants followed the link provided, informed consent was obtained using the online survey system through a series of approved questions. Participants were informed that the data would be converted to an anonymized format and that the data collected might be subject to publication. After completing the survey, participants received a written debrief through the online survey tool, were thanked for their participation, and were dismissed.

## Data Analysis

To test the hypotheses (see subsection 'Hypotheses'), we applied statistical methods to the collected survey data. To begin, descriptive statistics were used to illustrate both the user profile of participants and their attitudes toward Bitcoin.

Next, we examined three crucial variables: awareness, actionability, and responsibility. To assess the correlation between actionability and awareness, considering their nominal nature, we conducted a Chi-Square test.

Responsibility for addressing Bitcoin's electricity consumption was assessed using a 5-point Likert scale that measured participants' attitudes toward the responsibility of six different actors: Bitcoin miners, Bitcoin users, Bitcoin developers, intergovernmental organizations, the legislature, and federal agencies/regulators. We divided the six different actors into two distinct groups: non-government actors and government actors. Based on this assessment, we computed the average Likert scores by analysing participants' responses on the 5-point Likert scale.

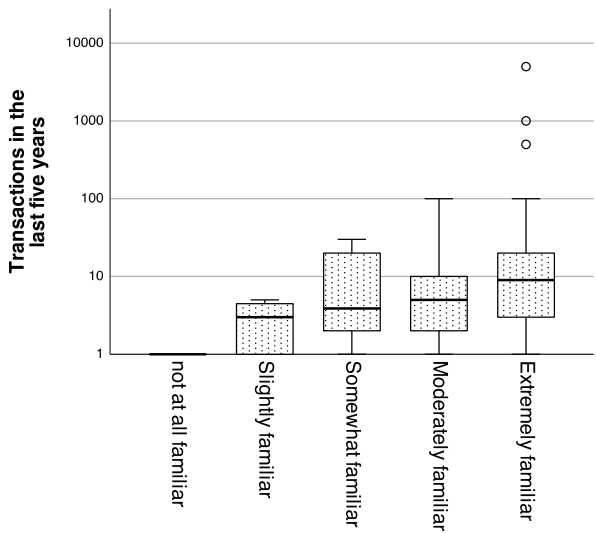

**Figure 7.** Respondents who report extreme familiarity with Bitcoin on average report the highest number of transactions (logarithmic scale). This group also includes outliers that report participation in very large numbers of transactions.

## Results

The results generated from the questionnaire provide insights into the environmental attitudes of the surveyed Bitcoin users and provide information on the key hypotheses (see subsection 'Hypotheses') in the areas of awareness (see subsection 'Awareness'), actionability (see subsection 'Actionability'), and responsibility (see subsection 'Responsibility').

### User Profiles

When analysing the user profile of participants (see Fig. 7), we found no relationship between transaction volume and familiarity with Bitcoin. We, however, observed some outliers that reported large numbers of Bitcoin transactions and self-reported being extremely familiar with this cryptocurrency. We found that the median number of transactions conducted in the last 5 years for all levels of experience was <10. When analysing how participants obtained Bitcoin (see Fig. 8), we found that online platforms were by far the most popular method, with 79.1% of participants having used them to acquire Bitcoin in the past. Few participants (2.5%) used dedicated kiosks (i.e. machines resembling cash machines) to acquire Bitcoin. The main motivation to acquire Bitcoin (reported by 40.5% of the participants) was as a long-term investment or retirement fund. Only 3.8% mentioned avoiding government regulation as a reason for obtaining Bitcoin[4].

In addition, we analyse the concerns that participants reported about the possible effects of climate change on their environment. Here, we found great consternation among participants with the mode of responses being extremely concerned for all the effects provided. Participants expressed significant concerns about freshwater resources, rising temperatures, and extreme weather events, although they were less troubled by variable rainfall. A total of 56.3% of participants believed that Bitcoin's electricity consumption contributes significantly to global $CO_2$ emissions, with almost all of these (93.3%) also believing that the $CO_2$ emissions caused by Bitcoin contribute to climate change. A total of 65.2% of overall participants felt that measures to reduce the $CO_2$ footprint of Bitcoin should be taken now. A minority of 42.7% of the participants who answered the relevant question supported the view that Bitcoin users should move away from Bitcoin to other cryptocurrencies in the interest of reducing $CO_2$ emissions. Some of these participants provided the names of alternative blockchain-based cryptocurrencies (e.g. Dogecoin and Ethereum). Others suggested alternative payment infrastructure tokens such as Ripple's XRP.

---

[4] This is likely under-reported because of a 'chilling effect': a condition in which prospective participants refrain from behaviour that deviates from the perceived rules, norms, and guidelines of a powerful supervisor for fear of negative consequences [167]: in this case, Nigerian authorities.

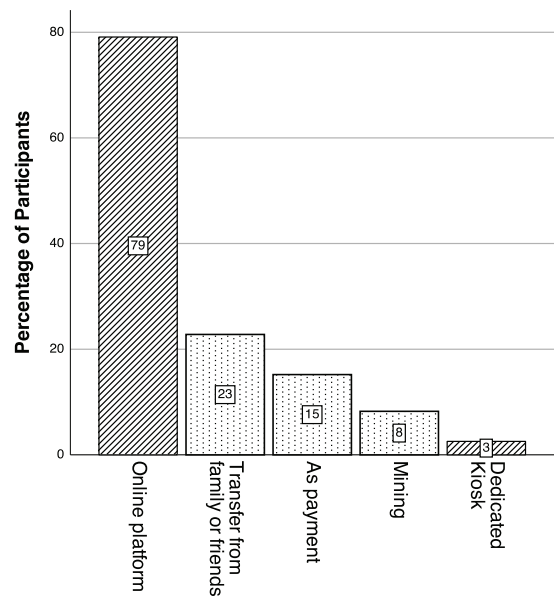

**Figure 8.** Participants have acquired Bitcoin through different channels. Online platforms and dedicated kiosks lend themselves well to digital energy labelling (see section 'Discussion') while others do not.

### Awareness

One of the key purposes of this questionnaire was to assess how realistic the estimates of the total Bitcoin electricity consumption made by the participants were. Here, we found that most participants (68.4%) significantly underestimated the energy demand of Bitcoin, whereas only a minority (16.5%) overestimated it (see Fig. 9). This goes beyond what is expected under random conditions: 50% of participants randomly selecting would underestimate electricity consumption, ∼16.7% would accurately assess it, and ∼33.3% would overestimate it.

### Actionability

We provided a variety of reference points to participants (see subsection 'Materials') to assess actionability. Based on the participants' estimates of the overall Bitcoin electricity consumption, we separated the participants into two groups: those who estimated energy demand correctly and those who did not. Thus, being supportive of measures and estimating the electricity consumption of Bitcoin correctly both constitute dichotomous variables.

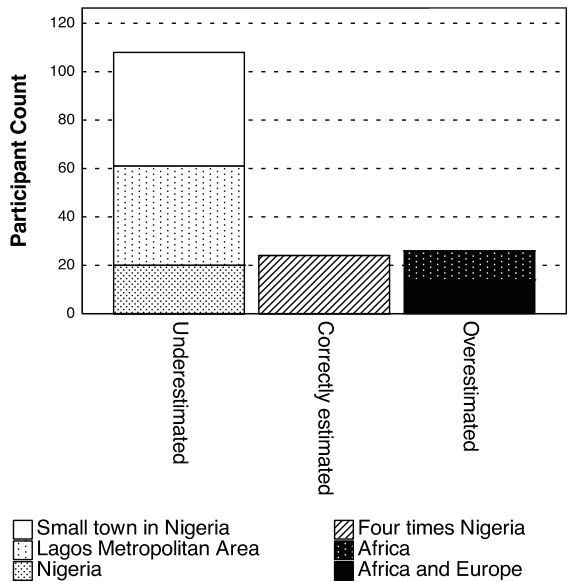

**Figure 9.** Most participants underestimate the overall energy demand of Bitcoin.

**Table 2.** We found a medium correlation between estimating the electricity consumption correctly and being supportive of measures

| Supportive | Estimates | | Total | $\chi^2$ |
| --- | --- | --- | --- | --- |
| | Correct | Incorrect | | |
| Yes | 83 | 20 | 103 | 4.105* |
| | 61.90% | 83.30% | 65.20% | |
| No | 51 | 4 | 55 | |
| | 38.10% | 16.70% | 34.80% | |
| Total | 134 | 23 | 158 | |

*Correlation is significant at the 0.05 level.

Subsequently, we conducted a Chi-Square test to analyse the relationship between supporting measures and estimating the electricity consumption of Bitcoin correctly: we found a medium correlation between these two variables ($\chi^2$ (1, $N = 158$) = 4.105, $p = 0.043 < 0.05$, $\phi_c = 0.16$, see Table 2). Specifically, a *post hoc* comparison test with correction showed that under $\alpha = 0.05$, the proportion of participants supporting measures in the correct estimates group (83.3%) is higher than in the incorrect estimates group (61.9%). Consequently, the proportion of the non-supporting measures group in the correct estimates group (16.7%) is lower than that of the incorrect estimates group (38.1%).

### Responsibility

Where participants did see the need for action, they felt that Bitcoin miners, Bitcoin users, and Bitcoin developers should be taking action instead of intergovernmental organizations, the legislature, and federal agencies or regulators (see Fig. 10).

To further examine participants' notion of the responsible actors, we rendered paired sample *t*-tests to compare the mean Likert scores in support of non-government actors (averaged by Bitcoin miners, Bitcoin users, and Bitcoin developers) and government actors (averaged by intergovernmental organizations, the legislature, and federal agencies or regulators). We observe

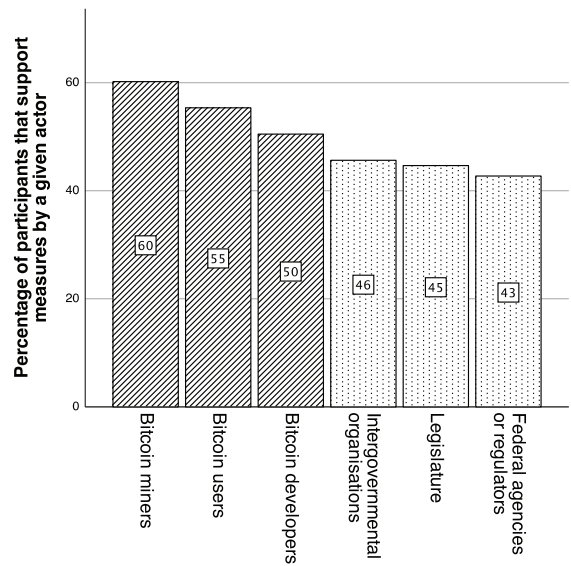

**Figure 10.** 60% of those participants that support measures in principle find that Bitcoin miners should act, while only 43% find that federal agencies or regulators should. A divide between non-governmental actors (top 3) and governmental actors (bottom 3) is noticeable.

**Table 3.** On average, participants hold private sector actors more accountable

| Group | $\mu$ | $\sigma$ | $t$ |
| --- | --- | --- | --- |
| Non-government actors | 3.49 | 1.02 | 2.943** |
| Government actors | 3.26 | 1.15 | |

$\mu$: Mean, $\sigma$: Standard deviation; ** Significant at the 0.01 level.

a significant difference in mean Likert scores in support of non-government actors and government actors ($t = 2.943$, $P = 0.004$, see Table 3). On average, participants expressed a stronger expectation of responsibility toward non-government actors ($M = 3.490$) compared with government actors ($M = 3.260$).

Some participants named alternative actors they felt were responsible: these included Bitcoin exchanges, wealthy individuals, and activists. Where participants did not feel that action to reduce the $CO_2$ footprint of Bitcoin should be taken now, they predominantly articulated two reasons for this perspective (both with 30.1%): they brought forward the view that future technological improvements would reduce Bitcoin's electricity demand and/or that the environmental impact of Bitcoin is acceptable for the benefit it provides. No meaningful alternative reasons were provided in the free-text fields.

## Discussion

Our results are largely corroborated by previous research. Specifically, they support the results of Eigbe [155], who previously pointed out gaps in the technical expertise of Nigerian Bitcoin users, as well as the findings of Steinmetz et al. [37] and Duggan [168], which yield similar conclusions, although outside of Nigeria. The results furthermore broadly align with consumer knowledge assessments in the broader financial products space, which showed that consumers often had little knowledge of the key properties of the products they were using [169, 170]. Although previous research has focused on technical or financial dimensions of user attitudes alone (see section 'Related Work'), the results of this study demonstrate that, throughout the user base, the concern over the effects of climate change is significant. These results should be taken into account when designing policies to respond to the high electricity consumption of cryptocurrencies.

To develop effective strategies to reduce the popularity of PoW cryptocurrencies, and therefore, ultimately, their electricity demand, decision-makers must first realize that such strategies cannot be targeted at miners alone. Although miners are, in fact, almost solely responsible for the energy footprint of cryptocurrencies (see section 'Cryptocurrency Mining'), they can quickly relocate their activities to other regions where there are fewer legal restrictions (see subsection 'The Chinese Example'). Relocating allows them to evade regulatory access without affecting the end users of the respective cryptocurrency because those are oblivious to where mining hardware is operated. A more effective strategy, instead, focuses on the end users of cryptocurrencies by empowering them to make more sustainable choices. This increase in transparency is a potential enabler for a consumer movement away from unsustainable cryptocurrencies. Such a consumer movement may result in a systematic reduction of the carbon footprint of unsustainable cryptocurrencies, beyond the individual user, should the expected price effects described earlier (see subsection 'Cryptocurrency Mining') materialize.

The finding that users who correctly assess the sustainability parameters of cryptocurrencies tend to show more support for measures indicates that consumer education is a promising tool for policymakers. Care must, however, be taken that cryptocurrencies are not portrayed in an all-encompassing and overly negative way: after all, our results do neither support nor rule out a correlation between 'overestimating' electricity consumption and supporting measures. Rather, policymakers should initiate measures that achieve basic consumer education and provide users of cryptocurrencies with a realistic view of their electricity consumption and economic parameters.

Energy labelling, i.e. providing key sustainability metrics to cryptocurrency users at the point of exchange, is one potentially suitable measure to achieve customer education. Such labels would allow users to compare the electricity consumption characteristics in this vast market, thereby allowing them to take sustainability into consideration when making cryptocurrency purchasing decisions. The concept of energy labelling aligns with the broader discussion of the importance of transparency and adequate disclosure in the blockchain and cryptocurrency industry [171]. Although little is known about the effectiveness of this intervention in the context of cryptocurrencies, the assessment of a protocol's consensus algorithm has previously been contemplated as a key environmental metric [172], and early research into measures to reduce the carbon impact of digital behaviours has produced promising results [173]. Furthermore, results from the field of household appliances, where energy labels are common, give cause for optimism: here, it was found that customers are aware of the information on labels [174] and comprehend it [175], albeit being confused by changes in labelling schemes [176]. Ultimately, consumers were found to make better decisions when guided by labels [177]. Furthermore, consumers were found to attach a value to energy efficiency beyond the prospect of reducing costs [178]. Even though sustainability awareness may differ between countries [179], it seems conceivable that energy labelling initiatives present an effective long-term energy efficiency policy for cryptocurrencies that may promote green innovation [180].

## Conclusions

The data obtained suggest that most Bitcoin users underestimate its electricity consumption. Our study also demonstrates a correlation between participants' ability to estimate the electricity consumption of Bitcoin correctly and their support of measures to counteract Bitcoin's $CO_2$ footprint. Furthermore, we find that users predominantly hold private actors (e.g. Bitcoin miners, users, and developers) responsible for addressing Bitcoin's energy demand. Subsequently, the empirical results lend support to all three hypotheses posited.

Taking into account the current trajectory of $CO_2$ emissions, regulators face unprecedented pressure to introduce policies to avoid a climate catastrophe. Cryptocurrencies based on PoW consume large amounts of electricity, while arguably providing very similar benefits to those built on alternative consensus mechanisms that are orders of magnitude less energy-demanding. The counteracting of the enormous electricity consumption of PoW-based cryptocurrencies must therefore be urgently attended to by policymakers, not least since cryptocurrencies are now ubiquitous and no longer exclusive to users with specific demographic or regional characteristics. Improving customer knowledge about cryptocurrency sustainability could lead to more sustainable consumer behaviour.

Therefore, in this work, we recommend a specific course of action to promote customer knowledge: confronting users with the consequences of their cryptocurrency choices through energy labelling. Although this proposal has not yet been tested, the key results of this work suggest that it may improve sustainability.

## Limitations and Future Work

It is important to note that our study is based on a small sample with a narrow scope because it focused solely on one asset and country and was obtained by convenience sampling. This sampling method may introduce biases from self-selection, inadvertent selection of specific groups, and recruitment channel preferences. Furthermore, the reliability of the data we collected is impacted by the challenging legal situation in Nigeria that may prevent cryptocurrency users from publicly acknowledging their activities. These factors warrant caution when generalizing our findings to larger populations.

In the future, experiments should evaluate the impact of presenting energy labels at the point of exchange[5] to test our policy suggestion. This will provide valuable insights into consumer behaviour. In addition, future research should focus on developing metrics that mitigate the misunderstandings around the concepts of network-wide and per-transaction electricity consumption measurements in Blockchain energy demand research, thereby creating criteria that are intuitive to experts and laypeople alike.

## Ethics Approval

Ethical clearance for this project was granted by the King's College London ethics committee under ethical clearance reference number *MRSP-21/22–27 025*. Freely given informed consent to participate in the study was obtained from all participants.

---

[5] Such research effort is undertaken by Drăgnoiu et al. [181].

## Availability of Data and Materials

The data underlying this article are available in Mendeley Data, at https://dx.doi.org/10.17632/j5j3gh4ps4.1. The questionnaire used in this study is available within the article's supplementary materials.

## Study Funding

The funds used for participant recruitment were sourced from a research grant by the University College London Centre for Blockchain Technologies, although these funds were not explicitly dedicated to this particular project.

## Competing Interests

M.P. reports a relationship with Amazon Web Services EMEA SARL, which includes employment. M.P. reports a relationship with R3 Ltd., which includes employment. M.P. reports a relationship with the University College London Centre for Blockchain Technologies, which includes consulting or advisory. S.O. reports a relationship with Stojeka Consulting Nig, Ltd., which includes consulting or advisory and equity or stocks. Stojeka Consulting Nig, Ltd. was commissioned as a research data collection provider. Competing interests were evaluated after the completion of the initial draft of this study in July 2022.

## Author Contributions Statement

Conceptualization: M.P. and S.O.; Data curation: M.P., S.O., and Z.W.; Formal analysis: Z.W.; Funding acquisition: M.P.; Investigation: M.P., S.O., A.D., O.E.I., F.P., J.S., and Z.W.; Methodology: M.P. and S.O.; Project administration: M.P. and S.O.; Supervision: M.P.; Visualisation: M.P.; Writing—original draft: M.P., S.O., A.D., O.E.I., F.P., J.S., and Z.W.; Writing—review & editing: M.P., S.O., A.D., O.E.I., F.P., J.S., and Z.W.

## Acknowledgements

We are deeply indebted to reviewer no. 4, whose unwavering passion for energy metrics fueled numerous revisions and caffeinated many late nights. Your tenacity, sir/madam, is laudable.

## References

1. Mikhaylov A. Cryptocurrency market analysis from the open innovation perspective. *Journal of Open Innovation: Technology, Market, and Complexity* 2020;**6**:197
2. Ante L. The non-fungible token (NFT) market and its relationship with Bitcoin and Ethereum. *FinTech* 2022;**1**:216–24
3. R. de Best. Market capitalization of Bitcoins. *Statista*, Nov. 2021. https://www.statista.com/statistics/377382. Last accessed: 24 November 2021.
4. Park S, Im S, Seol Y *et al.* Nodes in the Bitcoin network: comparative measurement study and survey. *IEEE Access* 2019;**7**:57009–22
5. Fang F, Ventre C, Basios M *et al.* Cryptocurrency trading: a comprehensive survey. *Financial Innovation* 2022;**8**.
6. Walsh M. Bitcoin, cryptocurrencies & the climate crisis. *Irish Marxist Review* 2021;**10**:80–9
7. Badea L, Mungiu-Pupazan MC. The economic and environmental impact of Bitcoin. *IEEE Access* 2021;**9**:48091–104
8. de Vries A. Bitcoin's growing energy problem. *Joule* 2018;**2**:801–5
9. Gehlot S, Dhall A. Evaluating the sustainability of Bitcoin. *Mathematical Statistician and Engineering Applications* 2022;**71**:139–51
10. Truby J. Decarbonizing Bitcoin: law and policy choices for reducing the energy consumption of blockchain technologies and digital currencies. *Energy Res Soc Sci* 2018;**44**:399–410
11. Gallersdorfer U, Klaaßen L, Stoll C. Energy consumption of cryptocurrencies beyond Bitcoin. *Joule* 2020;**4**:1843–6
12. Lei N, Masanet E, Koomey J. Best practices for analyzing the direct energy use of blockchain technology systems: review and policy recommendations. *Energy Policy* 2021;**156**:112422
13. de Vries A, Gallersdorfer U, Klaaßen L *et al.* Revisiting Bitcoin's carbon footprint. *Joule* 2022;**6**:498–502
14. Miraz MH, Excell PS, Sobayel K. Evaluation of green alternatives for blockchain proof-of-work (PoW) approach. *Annals of Emerging Technologies in Computing* 2021;**5**:54–9
15. de Vries A. Cryptocurrencies on the road to sustainability: ethereum paving the way for Bitcoin. *Patterns* 2022;**4**:100633
16. Kethineni S, Cao Y. The rise in popularity of cryptocurrency and associated criminal activity. *International Criminal Justice Review* 2020;**30**:325–44
17. Treiblmaier H, Gorbunov E. On the malleability of consumer attitudes toward disruptive technologies: a pilot study of cryptocurrencies. *Information* 2022;**13**:
18. Brown-Cohen J, Narayanan A, Psomas A *et al.* Formal barriers to longest-chain proof-of-stake protocols. In: *Proceedings of the 2019 Conference on Economics and Computation.* Phoenix, AZ, USA: ACM, 2019, 459–73.
19. Houy N. It will cost you nothing to 'kill' a proof-of-stake cryptocurrency. *Econ Bull* 2014;**34**:1038–44
20. Shifferaw Y, Lemma S. Limitations of proof of stake algorithm in blockchain: a review. *Zede Journal of Ethiopian Engineers and Architects* 2021;**39**:81–95
21. Kiayias A, Russell A, David B *et al.* Ouroboros: a provably secure proof-of-stake blockchain protocol. In: Katz J, Shacham H, (eds.), *Proceedings of the 37th Annual International Cryptology Conference, volume 10401 of Lecture Notes in Computer Science.* Santa Barbara, CA, USA: Springer, 2017, 357–88
22. Rieger A, Roth T, Sedlmeir J *et al.* We need a broader debate on the sustainability of blockchain. *Joule* 2022;**6**:1137–41
23. Saleh E. Blockchain without waste: proof-of-stake. *Rev Financ Stud* 2020;**34**:1156–90
24. Nair PR, Dorai DR Evaluation of performance and security of proof of work and proof of stake using blockchain. In: *Proceedings of the 3rd International Conference on Intelligent Communication Technologies and Virtual Mobile Networks.* Tirunelveli, India: IEEE, 2021, 279–83.
25. Sedlmeir J, Buhl HU, Fridgen G *et al.* The energy consumption of blockchain technology: beyond myth. *Bus Inf Syst Eng* 2020;**62**:599–608
26. Agur I, Deodoro J, Lavayssiere X *et al.* Digital currencies and energy consumption. *FinTech Notes 2022/006*, International Monetary Fund, Washington, DC, USA, 2022. https://www.imf.org/en/Publications/fintech-notes/Issues/2022/06/07/Digital-Currencies-and-Energy-Consumption-517866. Last accessed: 12 June 2022
27. Agur I, Lavayssiere X, Bauer GV *et al.* Lessons from crypto assets for the design of energy efficient digital currencies. *Ecol Econ* 2023;**212**:107888
28. Ioannou I. Legal regulation of virtual currencies: illicit activities and current developments in the realm of payment systems. *The King's Student Law Review* 2020;**11**:25–52

29. Truby J, Brown RD, Dahdal A *et al.* Blockchain, climate damage, and death: policy interventions to reduce the carbon emissions, mortality, and net-zero implications of non-fungible tokens and Bitcoin. *Energy Res Soc Sci* 2022;**88**:102499
30. Yusuf IA, Salaudeen MB, Ogbuji IA. Exchange rate fluctuation and inflation nexus in Nigeria: the case of recent recession. *Journal of Economic Impact* 2022;**4**:81–7
31. Osuagwu ES, Isola WA, Nwaogwugwu IC. Measuring technical efficiency and productivity change in the Nigerian banking sector: a comparison of non-parametric and parametric techniques. *Afr Dev Rev* 2018;**30**:490–501
32. T. Lawal. How Nigerians are increasingly turning to BTC and other altcoins. *Cryptotvplus*, 2021. https://cryptotvplus.com/2021/03/how-nigerians-areincreasingly-turning-to-btc-and-other-altcoins/. Last Accessed: 10 July 2022.
33. Bakare A. Cryptocurrency prohibition in Nigerians' best interest – Emefiele. *CBN Update* 2021;**3**:16–7
34. R. Auer and D. Tercero-Lucas. *Distrust or Speculation? The Socioeconomic Drivers of US Cryptocurrency Investments*. Working Paper 951, Bank for International Settlements, Basel, Switzerland, 2021. https://www.bis.org/publ/work951.pdf. Last Accessed: 10 July 2022
35. Baek C, Elbeck M. Bitcoins as an investment or speculative vehicle? A first look. *Appl Econ Lett* 2015;**22**:30–4
36. Glaser F, Zimmermann K, Haferkorn M *et al.* Bitcoin – asset or currency? Revealing users' hidden intentions. In: *Proceedings of the 22nd European Conference on Information Systems*. Tel Aviv, Israel: AIS, 2014.
37. Steinmetz F, von Meduna M, Ante L *et al.* Ownership, uses and perceptions of cryptocurrency: results from a population survey. *Technol Forecast Soc Chang* 2021;**173**:121073
38. Abah RC. Rural perception to the effects of climate change in Otukpo, Nigeria. *Journal of Agriculture and Environment for International Development* 2014;**108**:153–66
39. H. Haider. *Climate Change in Nigeria: Impacts and Responses*. K4D Helpdesk Report 675, Institute of Development Studies, Brighton, UK, 2019. https://opendocs.ids.ac.uk/opendocs/bitstream/handle/20.500.12413/14761/675_Climate_Change_in_Nigeria.pdf. Last accessed: 28 November 2021.
40. Mustapha B, Salau ES, Galadima OE *et al.* Knowledge, perception and adaptation strategies to climate change among farmers of central state Nigeria. *Sustainable Agriculture Research* 2013;**2**:107–17
41. Ku-Mahamud KR, Omar M, Bakar NAA *et al.* Awareness, trust, and adoption of blockchain technology and cryptocurrency among blockchain communities in Malaysia. *International Journal on Advanced Science, Engineering and Information Technology* 2019;**9**:1217
42. Shehhi AA, Oudah M, Aung Z Investigating factors behind choosing a cryptocurrency. In: *Proceedings of the 2014 International Conference on Industrial Engineering and Engineering Management*. Selangor, Malaysia: IEEE, 2014, 1443–7.
43. Bord RJ, OConnor RE, Fisher A. In what sense does the public need to understand global climate change? *Public Underst Sci* 2000;**9**:205–18
44. Bratspies RM. Cryptocurrency and the myth of the trustless transaction. *Michigan Technology Law Review* 2018;**25**:1
45. S. Nakamoto. *Bitcoin: A Peer-to-Peer Electronic Cash System*, 2008. https://bitcoin.org/bitcoin.pdf. Last Accessed: 21 May 2022.
46. Garriga M, Palma SD, Arias M *et al.* Blockchain and cryptocurrencies: a classification and comparison of architecture drivers. *Concurrency and Computation: Practice and Experience* 2020;**33**:e5992
47. Butijn B-J, Tamburri DA, van den Heuvel W-J. Blockchains. *ACM Comput Surv* 2021;**53**:1–37
48. Zhang Y Blockchain. In: Shen X, Lin X, Zhang K, (eds.), *Encyclopedia of Wireless Networks*. Cham, Switzerland: Springer, 2020, 115–8
49. Gregoriadis M, Muth R, Florian M Analysis of arbitrary content on blockchain-based systems using BigQuery. In: *Companion Proceedings of the Web Conference 2022*. Lyon, France: ACM, 2022, 478–87
50. Wu J, Liu J, Zhao Y *et al.* Analysis of cryptocurrency transactions from a network perspective: an overview. *J Netw Comput Appl* 2021;**190**:103139
51. Serena L, Ferretti S, D'Angelo G. Cryptocurrencies activity as a complex network: analysis of transactions graphs. *Peer-to-Peer Networking and Applications* 2021;**15**:839–53
52. Tai S, Eberhardt J, Klems M Not ACID, not BASE, but SALT. In: Ferguson D, Munoz VM, Cardoso J *et al.*, (eds.), *Proceedings of the 7th International Conference on Cloud Computing and Services Science*. Porto, Portugal: ACM, SciTePress, 2017, 755–64
53. Mukhopadhyay U, Skjellum A, Hambolu O *et al.* A brief survey of cryptocurrency systems. In: *Proceedings of the 14th Annual Conference on Privacy, Security and Trust*. Auckland, New Zealand: IEEE, 2016
54. Pontiveros BBF, Norvill R, State R Monitoring the transaction selection policy of Bitcoin mining pools. In: *Proceedings of the 2018 Symposium on Network Operations and Management*. Taipei, Taiwan: IEEE, 2018, 1–6s
55. Ehrenberg AJ, King JL. Blockchain in context. *Inf Syst Front* 2020;**22**:29–35
56. Platt M, McBurney P. Sybil in the haystack: a comprehensive review of blockchain consensus mechanisms in search of strong Sybil attack resistance. *Algorithms* 2023;**16**:34
57. Wang W, Hoang DT, Hu P *et al.* A survey on consensus mechanisms and mining strategy management in blockchain networks. *IEEE Access* 2019;**7**:22328–70
58. Lamport L. The part-time parliament. *ACM Trans Comput Syst* 1998;**16**:133–69
59. Castro M, Liskov B Practical Byzantine fault tolerance. In: *Proceedings of the 3rd Symposium on Operating Systems Design and Implementation*. New Orleans, LA, USA: ACM, 1999, 173–86
60. Douceur JR The Sybil attack. In: Druschel P, Kaashoek F, Rowstron A, (eds.), *Proceedings of the 1st International Workshop on Peer-to-Peer Systems, volume 2429 of Lecture Notes in Computer Science*. Cambridge, MA, USA: Springer, 2002, 251–60
61. Dwork C, Naor M Pricing via processing or combatting junk mail. In: Brickell EF, (ed.), *Proceedings of the 12th Annual International Cryptology Conference, volume 740 of Lecture Notes in Computer Science*. Santa Barbara, CA, USA: Springer, 1992, 139–47
62. A. Back. Hashcash – a denial of service counter-measure, 2002. http://www.hashcash.org/hashcash.pdf. Last Accessed: 10 July 2022.
63. de Vries A. Bitcoin boom: what rising prices mean for the network's energy consumption. *Joule* 2021;**5**:509–13
64. Sedlmeir J, Buhl HU, Fridgen G *et al.* Ein Blick auf aktuelle Entwicklungen bei Blockchains und deren Auswirkungen auf den Energieverbrauch. *Informatik Spektrum* 2020b;**43**:391–404
65. Stinner J On the economics of Bitcoin mining: a theoretical framework and simulation evidence. In: *Proceedings of the 43rd International Conference on Information Systems*. Copenhagen, Denmark: AIS, 2022
66. Kohli V, Chakravarty S, Chamola V *et al.* An analysis of energy consumption and carbon footprints of cryptocurrencies and

possible solutions. *Digital Communications and Networks* 2023;**9**: 79–89
67. Maiti M, Vukovic DB, M. Frömmel. Quantifying the asymmetric information flow between bitcoin prices and electricity consumption. *Financ Res Lett* 2023;**57**:104163
68. Zhang D, Chen XH, Lau CKM *et al.* Implications of cryptocurrency energy usage on climate change. *Technol Forecast Soc Chang* 2023;**187**:122219
69. Xie J, Yu FR, Huang T *et al.* A survey on the scalability of blockchain systems. *IEEE Netw* 2019;**33**:166–73
70. E. Georgiadis. How many transactions per second can Bitcoin really handle? Theoretically. 2019. https://eprint.iacr.org/2019/416. Last Accessed: 14 July 2022.
71. N. Carter. How much energy does Bitcoin actually consume? *Harv Bus Rev*, 2021. https://hbr.org/2021/05/how-much-energy-does-bitcoinactually-consume. Last Accessed: 15 January 2023.
72. Dittmar L, Praktiknjo A. Could Bitcoin emissions push global warming above 2C? *Nat Clim Chang* 2019;**9**:656–7
73. S. Bhambhwani, S. Delikouras, and G. Korniotis. *Blockchain Characteristics and the Cross-Section of Cryptocurrency Returns*. Discussion Paper 13724, Centre for Economic Policy Research, London, UK, 2019. https://cepr.org/publications/dp13724-6. Last Accessed: 4 February 2023.
74. Platt M, Sedlmeir J, Platt D *et al.* The energy footprint of blockchain consensus mechanisms beyond proof-of-work. In: *Companion Proceedings of the 21st International Conference on Software Quality, Reliability and Security*. Hainan, China: IEEE, 2021, 1135–44
75. Keenan TP Alice in blockchains: Surprising security pitfalls in PoW and PoS blockchain systems. In: *Proceedings of the 15th Annual Conference on Privacy, Security and Trust*. Calgary, AB, Canada: IEEE, 2017, 400–2
76. Ouyang Z, Shao J, Zeng Y PoW and PoS and related applications. In: *Proceedings of the 2021 International Conference on Electronic Information Engineering and Computer Science*. Changchun, China: IEEE, 2021, 59–62
77. Shin D, Bianco WT. In blockchain we trust: does blockchain itself generate trust? *Soc Sci Q* 2020;**101**:2522–38
78. Jang H, Han SH, Kim JH *et al.* Usability evaluation for cryptocurrency exchange. In: Gutierrez AMJ, Goonetilleke RS, Robielos RAC, (eds.), *Proceedings of the 2020 Joint Conference of the Asian Council on Ergonomics and Design and the Southeast Asian Network of Ergonomics Societies*. Virtual Event: Springer, 2021, 192–6
79. R. de Best. Bitcoin market dominance. *Statista*, June 2022. https://www.statista.com/statistics/1269669. Last Accessed: 4 July 2022.
80. Arslanian H Crypto exchanges. In: *The Book of Crypto*. Cham, Switzerland: Springer, 2022, 335–50
81. Zhou Z, Shen B Toward understanding the use of centralized exchanges for decentralized cryptocurrency. In: *Proceedings of the 13th International Conference on Applied Human Factors and Ergonomics*. New York, NY, USA, 2022
82. Hileman G The Bitcoin market potential index. In: Brenner M, Christin N, Johnson B *et al.*, (eds.), *Proceedings of the 19th International Conference on Financial Cryptography and Data Security*, volume 8976 of Lecture Notes in Computer Science. Carolina, Puerto Rico: Springer, 2015, 92–3
83. Jutel O. Blockchain financialization, neo-colonialism, and Binance. *Frontiers in Blockchain* 2023;**6**:1160257
84. O. Adesina. Nigeria attracts more Bitcoin interest than any country globally. *Nairametrics*, 2020. https://nairametrics.com/2020/08/08/nigeria-attractsmore-bitcoin-interest-than-any-country-globally/. Last Accessed: 10 July 2022.
85. Olubusoye OE, Salisu AA, Olofin SO. Youth unemployment in Nigeria: nature, causes and solutions. *Qual Quant* 2022;**57**: 1125–57
86. Burns S The Silicon Savannah: exploring the promise of cryptocurrency in Africa. In: Caton J, (ed.), *The Economics of Blockchain and Cryptocurrency*. Edward Elgar Publishing, 2022, 69–94
87. Zhao J. Do economic crises cause trading in Bitcoin? *Review of Behavioral Finance* 2022;**14**:465–90
88. BBC News. Cryptocurrencies: *Why Nigeria is a Global Leader in Bitcoin Trade*, Feb. 2021. https://www.bbc.com/news/world-africa-56169917. Last Accessed: 10 July 2022.
89. Onyekwere E, Ogwueleka FN, Irhebhude ME. Adoption and sustainability of bitcoin and the blockchain technology in Nigeria. *Int J Inf Technol* 2023;**15**:2793–804
90. Ediagbonya V, Tioluwani TC The growth and regulatory challenges of cryptocurrency transactions in Nigeria. In: *The Complexities of Sustainability*. Singapore: World Scientific, 2022, 267–97
91. Emmanuel OT, Michael AA. Forensic accounting: breaking the nexus between financial cybercrime and terrorist financing in Nigeria. *Journal of Auditing, Finance, and Forensic Accounting* 2020;**8**:55–66
92. Kothari N Cryptocurrency fraud: a glance into the perimeter of fraud. In: *Financial Crimes*. Cham, Switzerland: Springer, 2023, 109–29
93. Ewuzie OC, Obioma NR, Gift UO *et al.* Youth unemployment and cybercrime in Nigeria. *African Renaissance* 2023;**20**:177–99
94. Ibrahim S. Social and contextual taxonomy of cybercrime: socioeconomic theory of Nigerian cybercriminals. *International Journal of Law, Crime and Justice* 2016;**47**:44–57
95. Lazarus S, Okolorie GU. The bifurcation of the Nigerian cybercriminals: narratives of the economic and financial crimes commission (EFCC) agents. *Telematics Inform* 2019;**40**: 14–26
96. Okosun O, Ilo U. The evolution of the Nigerian prince scam. *Journal of Financial Crime* 2022; Vol. ahead-of-print No. ahead-of-print
97. Butler S Cyber 9/11 will not take place: a user perspective of Bitcoin and cryptocurrencies from underground and dark net forums. In: Groß T, Vigano L, (eds.), *Proceedings of the 10th International Workshop on Socio-Technical Aspects in Security and Trust*. Virtual Event: Springer, 2021, 135–53
98. Stringham EP. Banking regulation got you down? The rise of fintech and cryptointermediation in Africa. *Public Choice* 2023
99. Bakare A. Cryptocurrencies: SEC supports CBN position. *CBN Update* 2021;**3**:10
100. Nwanisobi O. Response to regulatory directive on cryptocurrencies. *CBN Update* 2021;**3**:7–9
101. Bakare A. Cryptocurrency trading: CBN orders banks to close operating accounts. *CBN Update* 2021;**3**:2
102. O. Adesina. Why CBN needs to lift crypto ban on Nigerian banks. *Nairametrics*, 2022. https://nairametrics.com/2022/03/11/why-cbn-needs-tolift-crypto-ban-on-nigerian-banks/. Last Accessed: 10 July 2022.
103. J. Uba. *Nigeria: Is Cryptocurrency Legal in Nigeria? – Actions Towards the Regulations of Cryptocurrency in Nigeria*. Mondaq, 2021. https://www.mondaq.com/nigeria/fin-tech/1105924. Last Accessed: 10 July 2022.
104. O. Adesina. Nigerian banks allegedly close accounts dealing with crypto. *Nairametrics*, 2021. https://nairametrics.

com/2021/02/11/nigerian-banksallegedly-close-accounts-dealing-with-crypto/. Last Accessed: 10 July 2022.
105. Olowodun D. eNaira: CBN appoints technical partner. *CBN Update* 2021;**3**:9
106. Ozili PK. Central bank digital currency research around the world: a review of literature. *Journal of Money Laundering Control* 2022;**26**:215–26
107. Chukwuere JE. The eNaira – opportunities and challenges. *Journal of Emerging Technologies* 2021;**1**:72–7
108. Ozili PK Redesigning the eNaira central bank digital currency (CBDC) for payments and macroeconomic effectiveness. In: Tyagi P, Grima S, Sood K et al., (eds.), *Smart Analytics, Artificial Intelligence and Sustainable Performance Management in a Global Digitalised Economy, volume 110B of Contemporary Studies in Economic and Financial Analysis*. Bingley, UK: Emerald Publishing, 2023, 189–97
109. J. Ree. *Nigeria's eNaira, One Year After*. Working Paper WP/23/104, International Monetary Fund, Washington, DC, USA, 2023. https://www.imf.org/en/Publications/WP/Issues/2023/05/16/Nigerias-eNaira-One-Year-After-533487
110. Ram AJ. Bitcoin as a new asset class. *Meditari Accountancy Research* 2019;**27**:147–68
111. van der Merwe A. A taxonomy of cryptocurrencies and other digital assets. *Rev Bus* 2021;**41**:30–43
112. Kulal A. Followness of altcoins in the dominance of Bitcoin: a phase analysis. *Macro Management & Public Policies* 2021;**3**:10–8
113. Grabowski M *Cryptocurrencies*. Routledge, London: UK, first edition, 2019
114. Basel Committee on Banking Supervision. Second consultation on the prudential treatment of cryptoasset exposures. *Consultative Document*, Bank for International Settlements, Basel, Switzerland, 2022. https://www.bis.org/bcbs/publ/d533.pdf. Last Accessed: 10 July 2022
115. Fletcher E, Larkin C, Corbet S. Countering money laundering and terrorist financing: a case for bitcoin regulation. *Res Int Bus Financ* 2021;**56**:101387
116. Sicignano GJ. Money laundering using cryptocurrency: the case of Bitcoin! *Athens Journal of Law* 2021;**7**:253–64
117. Jiang S, Li Y, Lu Q et al. Policy assessments for the carbon emission flows and sustainability of Bitcoin blockchain operation in China. *Nat Commun* 2021;**12**:1938
118. Wanat E. Are crypto-assets green enough? – An analysis of draft EU regulation on markets in crypto assets from the perspective of the European Green Deal. *Osteuropa Recht* 2021;**67**:237–50
119. Ahern D. Regulatory lag, regulatory friction and regulatory transition as fintech disenablers: calibrating an EU response to the regulatory sandbox phenomenon. *European Business Organization Law Review* 2021;**22**:395–432
120. Filippi PD, Mannan M, Reijers W. The alegality of blockchain technology. *Polic Soc* 2022;**41**:358–72
121. Ferreira A, Sandner P. Eu search for regulatory answers to crypto assets and their place in the financial markets' infrastructure. *Computer Law & Security Review* 2021;**43**:105632
122. Shin D, Ibahrine M. The socio-technical assemblages of blockchain system: how blockchains are framed and how the framing reflects societal contexts. *Digital Policy, Regulation and Governance* 2020;**22**:245–63
123. Shin D, Rice J. Cryptocurrency: a panacea for economic growth and sustainability? A critical review of crypto innovation. *Telematics Inform* 2022;**71**:101830
124. Shin D, Stylos N, Asim M et al. How does the blockchain find its way in the UAE? The blockchain as a sociotechnical system. *Int J Technol Manag* 2022;**90**:122
125. Hammond S, Ehret T. *Cryptocurrency Regulations by Country*. Technical report, Thomson Reuters Institute, New York, NY, USA, 2022. https://www.thomsonreuters.com/en-us/posts/wpcontent/uploads/sites/20/2022/04/Cryptos-Report-Compendium-2022.pdf. Last accessed: 4 July 2022.
126. Silva EC, da Silva MM. Research contributions and challenges in DLT-based cryptocurrency regulation: a systematic mapping study. *Journal of Banking and Financial Technology* 2022;**6**: 63–82
127. United States Department of the Treasury. *General Explanations of the Administration's Fiscal Year 2024 Revenue Proposals. General Explanations of the Administration's Revenue Proposals*, United States Department of the Treasury, Washington, DC, USA, 2023. https://home.treasury.gov/system/files/131/General-Explanations-FY2024.pdf. Last accessed: 11 March 2023.
128. Oghan V. Environmental policies for green cryptocurrency mining. In: Akkaya S, Erguder B, (eds.), *Handbook of Research on Challenges in Public Economics in the Era of Globalization*, chapter 13. IGI Global, 2022, 217–27
129. Patel SK, Jhalani P. Formulation of variables of environmental taxation: a bibliometric analysis of scopus database (2001–2022). *Environ Dev Sustain* 2023;
130. C. Gola and J. Sedlmeir. *Addressing the Sustainability of Distributed Ledger Technology*. Occasional Paper 670, Bank of Italy, Rome, Italy, 2022. https://www.bancaditalia.it/pubblicazioni/qef/2022-0670/. Last accessed: 7 July 2022, Addressing the Sustainability of Distributed Ledger Technology.
131. Mathews R, Khan S. Balancing the benefits and burdens of blockchain on the environment: a barrier or boon for sustainability? *Environmental Law & Management* 2019;**31**:163–8
132. Luther WJ, Smith SS. Is Bitcoin a decentralized payment mechanism? *J Inst Econ* 2020;**16**:433–44
133. Fakunmoju SK, Banmore O, Gbadamosi A et al. Effect of cryptocurrency trading and monetary corrupt practices on Nigerian economic performance. *Binus Business Review* 2022;**13**: 31–40
134. Millard C. Blockchain and law: incompatible codes? *Computer Law & Security Review* 2018;**34**:843–6
135. Mezquita Y, Perez D, Gonzalez-Briones A et al. Cryptocurrencies, survey on legal frameworks and regulation around the world. In: Prieto J, Martınez FLB, Ferretti S et al., (eds.), *Proceedings of the 4th International Congress on Blockchain and Applications, volume 595 of Lecture Notes in Networks and Systems*. L'Aquila, Italy: Springer, 2023, 58–66
136. Okorie DI, Lin B. Did China's ICO ban alter the Bitcoin market? *International Review of Economics & Finance* 2020;**69**:977–93
137. Borri N, Shakhnov K. Regulation spillovers across cryptocurrency markets. *Financ Res Lett* 2020;**36**:101333
138. Chen C, Liu L. How effective is China's cryptocurrency trading ban? *Financ Res Lett* 2022;**46**:102429
139. Feinstein BD, Werbach K. The impact of cryptocurrency regulation on trading markets. *Journal of Financial Regulation* 2021;**7**: 48–99
140. Mikki S. Google Scholar compared to Web of Science. a literature review. *Nordic Journal of Information Literacy in Higher Education* 2009;**1**:
141. Ajzen I. The theory of planned behavior. *Organ Behav Hum Decis Process* 1991;**50**:179–211
142. Alaklabi S, Kang K Factors driving individuals' behavioural intention to adopt cryptocurrency. In: *Proceedings of the 34th International Business Information Management Association Conference*. Madrid, Spain: IBIMA, 2019, 2070–8

143. Albayati H, Kim SK, Rho JJ. Accepting financial transactions using blockchain technology and cryptocurrency: a customer perspective approach. *Technol Soc* 2020;**62**: 101320
144. Anser MK, Zaigham GHK, Rasheed MI et al. Social media usage and individuals' intentions toward adopting Bitcoin: the role of the theory of planned behavior and perceived risk. *Int J Commun Syst* 2020;**33**:e4590
145. Mazambani L, Mutambara E. Predicting fintech innovation adoption in South Africa: the case of cryptocurrency. *Afr J Econ Manag Stud* 2019;**11**:30–50
146. Schaupp LC, Festa M Cryptocurrency adoption and the road to regulation. In: *Proceedings of the 19th Annual International Conference on Digital Government Research*. Delft, Netherlands: ACM, 2018, 1–9
147. Pham QT, Phan HH, Cristofaro M et al. Examining the intention to invest in cryptocurrencies. *International Journal of Applied Behavioral Economics* 2021;**10**:59–79
148. Smutny Z, Sulc Z, Lansky J. Motivations, barriers and risk-taking when investing in cryptocurrencies. *Mathematics* 2021;**9**:1655
149. Kim M. A psychological approach to Bitcoin usage behavior in the era of COVID-19: focusing on the role of attitudes toward money. *J Retail Consum Serv* 2021;**62**:102606
150. Salcedo E, Gupta M. The effects of individual-level espoused national cultural values on the willingness to use Bitcoin-like blockchain currencies. *Int J Inf Manag* 2021;**60**: 102388
151. Bashir M, Strickland B, Bohr J What motivates people to use Bitcoin? In: Spiro E, Ahn Y-Y, (eds.), *Proceedings of the 8th International Conference on Social Informatics, volume 10047 of Lecture Notes in Computer Science*. Bellevue, WA, USA: Springer, 2016, 347–67
152. Gagarina M, Nestik T, Drobysheva T. Social and psychological predictors of youths' attitudes to cryptocurrency. *Behavioral Sciences* 2019;**9**:118
153. Dodd N. The social life of Bitcoin. *Theory, Culture & Society* 2017;**35**:35–56
154. Nnabuife SO, Jarrar Y. Online media coverage of BitCoin cryptocurrency in Nigeria: a study of selected online version of leading mainstream newspapers in Nigeria. *H-ermes Journal of Communication* 2018;**12**:141–72
155. Eigbe OE. Investigating the levels of awareness and adoption of digital currency in Nigeria: a case study of bitcoin. *The Information Technologist* 2018;**15**:75–82
156. Salawu MK, Moloi T. Benefits of legislating cryptocurrencies: perception of Nigerian professional accountants. *Academy of Accounting and Financial Studies Journal* 2018;**22**: 1–17
157. Jimoh SO, Benjamin OO. The effect of cryptocurrency returns volatility on stock prices and exchange rate returns volatility in Nigeria. *Acta Universitatis Danubius Œconomica* 2020;**16**: 200–13
158. Egbo OP, Ezeaku HC. Overview of the intermediary role of banks: the threat of cryptocurrency. *European Journal of Management and Marketing Studies* 2016;**1**:89–101
159. Aberu F, Ogede JS, Ewarawon JO. Black market exchange rate movement in Nigeria: does cryptocurrency matters? *Studies of Applied Economics* 2023;**41**
160. Gidigbi MO, Babarinde GF, Adewusi OA et al. Legal issues and risks associated with cryptocurrency (in Nigeria): regulation for use and integration for its possibilities. *International Journal of Blockchains and Cryptocurrencies* 2021;**2**:320
161. Ukwueze F. Cryptocurrency: towards regulating the unruly enigma of fintech in Nigeria and South Africa. *Potchefstroom Electronic Law Journal* 2021;**24**:1–38
162. Smuts N. What drives cryptocurrency prices? *ACM SIGMETRICS Performance Evaluation Review* 2019;**46**:131–4
163. Senkardes BG, Akadur O. A research on the factors affecting cryptocurrency investments within the gender context. *Journal of Business, Economics and Finance* 2021;**10**:178–89
164. Steinmetz F. The interrelations of cryptocurrency and gambling: results from a representative survey. *Comput Hum Behav* 2023;**138**:107437
165. Mora C, Rollins RL, Taladay K et al. Bitcoin emissions alone could push global warming above 2C. *Nat Clim Chang* 2018;**8**: 931–3
166. OECD. *OECD Consumer Insights Survey on Cryptoassets*. Report, Organisation for Economic Co-Operation and Development, Paris, France, 2019. https://www.oecd.org/financial/education/consumerinsights-survey-on-cryptoassets.pdf. Last accessed: 26 November 2021.
167. Schüll A. Das stell ich lieber nicht ins netz!-zum chilling effect und seinen konsequenzen. In: Gadatsch A, Ihne H, Monhemius J et al., (eds.), *Nachhaltiges Wirtschaften im digitalen Zeitalter*. Springer, 2018, 53–62
168. W. Duggan. Survey: 84% of Americans don't believe that Bitcoin investments are a threat to the environment. *Forbes Advisor*, 2022. https://www.forbes.com/advisor/investing/cryptocurrency/survey-84-dont-believethat-bitcoin-investments-threat-to-environment/. Last Accessed: 11 March 2023.
169. Ramchander M. Measuring consumer knowledge of life insurance products in South Africa. *South Afr J Bus Manag* 2016;**47**: 67–74
170. Sukumaran S, Bee TS, Wasiuzzaman S. Cryptocurrency as an investment: the Malaysian context. *Risks* 2022;**10**:86
171. Liebau D, Krapels NJ, Youngblom R. An exploratory essay on minimum disclosure requirements for cryptocurrency and utility token issuers. *Cryptoeconomic Systems* 2021;**1**
172. D. Liebau. Sustainability in the Web 3.0: how to assess ESG characteristics on smart-contract blockchain ecosystems, 2021. https://ssrn.com/abstract=3929939. Last Accessed: 19 August 2023.
173. Seger BT, Burkhardt J, Straub F et al. Reducing the individual carbon impact of video streaming: a seven-week intervention using information, goal setting, and feedback. *J Consum Policy* 2023;**46**:137–53
174. Waechter S, Sutterlin B, Siegrist M. Desired and undesired effects of energy labels – an eye-tracking study. *PLoS One* 2015;**10**:e0134132
175. Jeong G, Kim Y. The effects of energy efficiency and environmental labels on appliance choice in South Korea. *Energy Efficiency* 2014;**8**:559–76
176. Stasiuk K, Maison D. The influence of new and old energy labels on consumer judgements and decisions about household appliances. *Energies* 2022;**15**:1260
177. Davis LW, Metcalf GE. Does better information lead to better choices? Evidence from energy-efficiency labels. *J Assoc Environ Resour Econ* 2016;**3**:589–625
178. Andor MA, Gerster A, Sommer S. Consumer inattention, heuristic thinking and the role of energy labels. *Energy J* 2020;**41**: 83–112
179. Schallehn F, Valogianni K. Sustainability awareness and smart meter privacy concerns: the cases of US and Germany. *Energy Policy* 2022;**161**:112756

180. Li Z, Zhang R, Zhu H. Environmental regulations, social networks and corporate green innovation: how do social networks influence the implementation of environmental pilot policies? *Environ Dev Sustain* 2022

181. Drăgnoiu A-E, Platt M, Wang Z *et al*. The more you know: energy labelling enables more sustainable cryptocurrency investments. In: *Proceedings of the 43rd International Conference on Distributed Computing Systems*. Hong Kong: IEEE. In Press, 2023